\documentclass[pre,aps,superscriptaddress,fleqn,twocolumn]{revtex4}
\usepackage{amsmath,amssymb,graphicx}
\usepackage{times,mathptm}
\makeatletter
\renewcommand{\@biblabel}[1]{\hspace*{2ex}}%
\makeatother

\begin{document}

\addtolength{\textheight}{2\baselineskip} 

\title{Looking at structure, stability, and evolution of proteins through the
       principal eigenvector of contact matrices and hydrophobicity profiles}

\author{Ugo~Bastolla}
\affiliation{Centro~de~Astrobiolog{\'\i}a~(INTA-CSIC),
             28850~Torrej\'on~de~Ardoz, Spain}

\author{Markus~Porto}
\affiliation{Institut~f\"ur~Festk\"orperphysik,
             Technische~Universit\"at~Darmstadt,
             Hochschulstr.~8,  64289~Darmstadt, Germany}

\author{H.~Eduardo~Roman}
\altaffiliation[Present address:]{Dipartimento~di~Fisica,
                Universit\`a~di~Milano~Bicocca, Piazza~della~Scienza~3,
                20126~Milano, Italy}
\affiliation{Dipartimento~di~Fisica and INFN, Universit\`a~di~Milano,
             Via~Celoria~16, 20133~Milano, Italy}

\author{Michele~Vendruscolo}
\affiliation{Department~of~Chemistry, University~of~Cambridge,
             Lensfield~Road, Cambridge CB2~1EW, UK}

\date{\today}

\begin{abstract}
We review and further develop an analytical model that describes how
thermodynamic constraints on the stability of the native state influence
protein evolution in a site-specific manner. To this end, we represent both
protein sequences and protein structures as vectors: Structures are represented
by the principal eigenvector (PE) of the protein contact matrix, a quantity
that resembles closely the effective connectivity of each site; Sequences are
represented through the ``interactivity'' of each amino acid type, using novel
parameters that are correlated with hydropathy scales. These interactivity
parameters are more strongly correlated than the other hydropathy scales that
we examine with: (1)~The change upon mutations of the unfolding free energy of
proteins with two-states thermodynamics; (2)~Genomic properties as the
genome-size and the genome-wide GC content; (3)~The main eigenvectors of the
substitution matrices. The evolutionary average of the interactivity vector
correlates very strongly with the PE of a protein structure. Using this result,
we derive an analytic expression for site-specific distributions of amino acids
across protein families in the form of Boltzmann distributions whose ``inverse
temperature'' is a function of the PE component. We show that our predictions
are in agreement with site-specific amino acid distributions obtained from the
Protein Data Bank, and we determine the mutational model that best fits the
observed site-specific amino acid distributions. Interestingly, the optimal
model almost minimizes the rate at which deleterious mutations are eliminated
by natural selection.
\end{abstract}

\maketitle


\section{Introduction}

The need to maintain the thermodynamic stability of the native state is an
important determinant of protein evolution. This requirement has different
effects on the various positions of the protein, depending on their structural
environment. Therefore it is crucial to consider the effect of site-specific
structural constraints in models of protein evolution, such as those used to
estimate phylogenetic distances from the comparison of protein sequences (Nei
and Kumar, 2000) or to reconstruct phylogenetic trees through Maximum
Likelihood methods (Felsenstein, 1981).

The first and simplest models of this kind assumes a Gamma distribution of
site-specific substitution rates (see Nei and Kumar, 2000). The Gamma
distribution is flexible enough to interpolate between broad and narrow
distributions, and its free parameter improves considerably the fit between
models of evolution and multiple sequence alignments. Subsequently,
site-specific amino acid frequencies (Halpern and Bruno, 1998) and
site-specific substitution matrices determined for different structural
classes (Li\`o and Goldman, 1998; Koshi and Goldstein, 1998; Koshi
et al., 1999; Thorne, 2000; Parisi and Echave, 2001; Fornasari et
al., 2002) were shown to further improve the reconstruction of phylogenetic
trees. In these works, site-specificity is either empirically derived from the
data or simulated with the help of some protein model. Following the latter
approach, we recently showed that a structural property, the principal
eigenvector (PE) of the contact matrix, is a strong predictor of
site-specific amino acid conservation in models of protein evolution with
stability constraints (Bastolla et al., 2004b).

We represent the amino acid sequence through a real-valued, site-specific
profile, using the interactivity parameters derived by Bastolla et al.\
(2004b). The interactivity scale expresses the strength of effective
interactions for each amino acid type, and its main component is the
hydropathy. The interactivity parameters correlate more strongly than other
hydropathy scales with thermodynamic quantities like the thermodynamic effect
of a mutation, with genomic quantities such as the genome size and the
genome-wide GC content in bacterial genomes, and with the main component of
substitution matrices, as we discuss in the present paper.

By using the interactivity parameters and the PE as a structural indicator,
we recently derived analytic expressions to predict site-specific amino acid
distributions (Porto et al., 2004b).
Amino acid distributions with a similar functional form have also been
used previously by other authors (Koshi and Goldstein, 1998;
Koshi et al.\ 1999; Dokholyan et al.\ 2001; 2002).
Our predictions are in very good agreement with
the distributions observed in the Protein Data Bank.
Here we consider simple mutational models at the nucleotide level, and we
show that they improve the fit between predicted and observed distributions.
Interestingly, the mutational model that best fits the observed data is one
for which deleterious mutations eliminated by negative selection appear with
almost the lowest rate.


\section{Materials and methods}

\subsection{Contact matrix and principal eigenvector}

We represent a protein structure as a $N\times N$ binary contact matrix
$C_{ij}$, where $N$ is the number of residues. The element $C_{ij}$ is
equal to one if amino acids $i$ and $j$ are in contact in the
native structure (i.e.\ at least one pair of their respective heavy atoms is
closer than $4.5 \, \mbox{\AA}$), and zero otherwise. The contact matrix is
symmetric, so it has $N$ real eigenvalues. The PE is the eigenvector
corresponding to the maximum eigenvalue and is indicated by ${\mathbf{c}}$.
This vector maximizes the quadratic form $\sum_{ij} C_{ij} \, c_i \, c_j$ with
the constraint $\sum_i c_i^2=1$. In this sense, $c_i$ can be interpreted as
the effective connectivity of position $i$, since positions with large $c_i$
are in contact with as many as possible positions $j$ with large $c_j$. All its
components have the same sign, which we choose by convention to be positive.
Moreover, if the contact matrix represents a single connected graph (as for
single-domain globular proteins), the information contained in the principal
eigenvector is sufficient to reconstruct the whole contact matrix (Porto et
al., 2004a). As we will review in this paper, the PE provides a convenient
vectorial representation of a protein structure to be used in evolutionary
studies.

\subsection{Free energy and interactivity parameters}

The effective free energy for a sequence ${\mathbf{A}}$ in configuration
${\mathbf{C}}$ is estimated through an effective contact free energy function
$E({\mathbf{A}},{\mathbf{C}})$,
\begin{equation}\label{eq:energy}
\frac{E({\mathbf{A}},{\mathbf{C}})}{k_{\mathrm{B}} T} =
 \sum_{i < j} C_{ij} \, U(A_i, A_j) \, ,
\end{equation}
where ${\mathbf{U}}$ is a $20\times 20$ symmetric matrix whose element $U(a,b)$
represents the effective interaction, in units of $k_{\mathrm{B}} T$, of amino
acids $a$ and $b$; we use the interaction matrix derived by Bastolla et al.\
(2001). For this interaction matrix the free energy function,
Eq.~(\ref{eq:energy}), is lower for the native structure than for decoys
generated by threading (Bastolla et al., 2001).

The effective energy function, Eq.~(\ref{eq:energy}), can be approximated
through the main component of its spectral decomposition
$H({\mathbf{A}},{\mathbf{C}})$,
\begin{equation}\label{eq:hydro}
\frac{H({\mathbf{A}},{\mathbf{C}})}{k_{\mathrm{B}} T} \equiv
\epsilon_1 \sum_{i < j} C_{ij} \, h(A_i) \, h(A_j)\, .
\end{equation}
where $\epsilon_1 < 0$ is the largest eigenvalue (in absolute value) of the
matrix $U(a,b)$ and $h(a)$ is the corresponding eigenvector. The latter is
mainly due to the contribution of hydrophobic interactions, as it is well known
(Casari et al., 1992; Li et al., 1997). We therefore call the $N$-dimensional
vector $h(A_i)$ the Hydrophobicity Profile (HP) of sequence ${\mathbf{A}}$
(Bastolla et al., 2004b). The 20 parameters $h(a)$ obtained from the principal
eigenvector of the interaction matrix are called \textit{interactivity}
parameters (IH). We will also use an optimized interactivity scale that was
obtained by Bastolla et al.\ (2004b) by optimizing the correlation between the
HP and the PE over a large set of proteins. Since the PE represents an
effective connectivity, we indicate the latter scale with the abbreviation CH.

\subsection{Hydropathy scales}

We consider eleven hydropathy scales, including the two interactivity scales
described above. They are: (1)~The KD hydropathy scale, derived to identify
trans-membrane helices using diverse experimental data (Kyte and Doolittle,
1982); (2)~The L76 hydropathy scale, which was derived by using experimental
data and theoretical calculations (Levitt, 1976); (3)~The R88 hydropathy
scale, which is based on the transfer of solutes from water to alkane
solvents (Roseman, 1988); (4)~The augmented Whilmey-White (WW)
hydropathy scale, derived to
improve recognition of trans-membrane helices (Jayasinghe et al., 2001);
(5)~The G98 classification of amino acids into polar, hydrophobic, and
amphiphylic classes, adopted by Gu et al.\ (1998) to investigate the
relationship between the hydrophobicity of a protein and the nucleotide
composition of the corresponding gene; (6)~The MP hydropathy scale, derived
from statistical properties of globular proteins (Manavalan and Ponnuswamy,
1978); (7)~The AV hydropathy scale derived by averaging 127 normalized
hydropathy scales published in the literature (Palliser and Parry, 2001);
(8)~The FP hydropathy scale, derived from the experimental measurement of
octanol/water partition coefficients (Fauchere and Pliska, 1983); (9)~The Z04
scale, also called buriability, proposed by Zhou and Zhou (2004); (10)~The
interaction scale IH, obtained from the main eigenvector of the interaction
matrix $U(a,b)$ used in this work (Bastolla et al., 2004b); (11)~The optimized
interactivity scale, or connectivity scale CH, which maximizes the correlation
with the principal eigenvectors of protein contact matrices for a non-redundant
set of Protein Data Bank (PDB) structures (Bastolla et al., 2004b). All these
scales have pairwise correlation coefficients ranging from a minimum of $0.68$
(between the KD and L76 scales) to a maximum of $0.95$ (between IH and CH
scales).

\subsection{SCN model of neutral evolution}

In the Structurally Constrained Neutral (SCN) model (Bastolla et al., 1999, 2002,
2003a, 2003b), starting from the protein sequence in the PDB, amino acid mutations
are randomly proposed and accepted according to a stability criterion based on
an effective model of protein folding (Bastolla et al., 2000, 2001). This
``structural'' approach has been pioneered by Schuster and co-workers, with a
series of studies of neutral networks of RNA secondary structures (Schuster et
al., 1994; Huynen et al., 1996; Fontana and Schuster, 1998) and it has been
applied to proteins by several groups (Gutin et al., 1995;
Bornberg-Bauer, 1997; Bornberg-Bauer and Chan, 1999; Babajide et al., 1997;
Govindarajan and Goldstein, 1998; Bussemaker et al., 1997; Tiana et al.,
1998; Mirny and Shakhnovich, 1998; Dokholyan and Shakhnovich, 2001;
Parisi and Echave, 2001).

For testing the stability of a protein conformation, we use two computational
parameters: (i)~The effective energy per residue,
$E({\mathbf{A}},{\mathbf{C}})/N$, Eq.~(\ref{eq:energy}), where $N$ is the
protein length. This quantity correlates with the folding free energy per
residue for a set of 18 small proteins that are folding with two-states
thermodynamics (correlation coefficient $0.91$; U.~Bastolla, unpublished
result); (ii)~The normalized energy gap $\alpha$, which characterizes fast
folding model sequences (Bastolla et al., 1998) with well correlated energy
landscapes (Bryngelson and Wolynes, 1987; Goldstein et al., 1992; Abkevich et
al., 1994; Gutin et al., 1995; Klimov and Thirumalai, 1996). In the SCN model,
a mutated sequence is considered thermodynamically stable if both computational
parameters are above predetermined thresholds (Bastolla et al., 2003a).

Simulations of the SCN model were performed for several protein folds. They
all show the same two features: (1)~After a very long evolutionary time,
the protein sequence looses all recognizable similarity with the starting
sequence, despite conserving the stability of its native fold. (2)~Structural
conservation varies across protein positions. The main structural determinant
of evolutionary conservation is the principal eigenvector of the contact
matrix.

\subsection{Site-specific distributions from the PDB}

We obtain site-specific amino acid distributions from the PDB as follows.
First, we select a non-redundant set of single-domain globular proteins,
i.e.\ sequences with more than 90\% homology with other chains in the PDB are
excluded. A non-redundant list of proteins is given in the file
\texttt{nrdb90}, available at the PDB web site. Qualitatively equivalent
results were obtained using the list selected with the method by Hobohm
and Sanders (1994) with the much lower threshold of 25\% sequence identity.
Globular proteins are selected through the condition that the
fraction of contacts per residue should be larger than a length-dependent
threshold, $N_{\mathrm{c}}/N > 3.5 + 7.8 N^{-1/3}$. This functional form
represents the scaling of the number of contacts in globular proteins as a
function of chain length (the factor $N^{-1/3}$ comes from the surface to
volume ratio), and the two parameters are chosen so as to eliminate outliers
with respect to the general trend, which are mainly non-globular structures.
Single-domain proteins are selected by imposing the normalized variance of
the PE components to be smaller than a threshold,
$\left(1-N \left< c \right>^2 \right)/\left( N \left< c \right>^2 \right)
<1.5$.
Multi-domain proteins have large PE components inside their main domains and
small components outside them. The PE components would be exactly zero
outside the main domain if the domains do not form contacts. Therefore,
multi-domain proteins are characterized by a larger normalized variance of
PE components with respect to single-domain ones. The threshold of $1.5$
eliminates most of the known multi-domain proteins and very few of the known
single-domain proteins (data not shown).

The proteins are divided into groups according to their length. A first group
comprises proteins with less than $100$ amino acids and contains about $8
\times 10^3$ sites. A second group comprises proteins of length between $101$
and $200$ amino acids and contains $5.2 \times 10^4$ sites. These two groups
are joined to improve the statistics in cases where we are not interested in
length dependence. A third group comprises proteins with length between $201$
and $400$ amino acids and contains $6 \times 10^5$ sites. Within each group,
sites are divided in structural classes depending on the value of their
normalized PE component, $c_i/\left< c \right>$, using a bin width of $0.1$.
For each structural class, the distribution of the $20$ types of amino acids
is obtained. We refer to this distribution with the symbol $\pi_{c_i/\left< c
\right>}^{\mathrm{PDB}}(a)$, where $a$ indicates one of the 20 types of
amino acids.

\subsection{Distance measures for amino acid distributions}

The Jensen-Shannon divergence of two given distributions $P(a)$ and $Q(a)$ is
defined as
\begin{equation}
\begin{split}
& D_{\mathrm{JS}}(P,Q) = \\
& \quad \frac{1}{2}\sum_a \left[
P(a) \, \log_2\left(\frac{P(a)}{R(a)}\right)+
Q(a) \, \log_2\left(\frac{Q(a)}{R(a)}\right) \right] ,
\end{split}
\end{equation}
where $R(a) = [P(a)+Q(a)]/2$. This measure is equivalent to the difference of
the entropy of the `average distribution' $R$ minus the average entropy of
distributions $P$ and $Q$. It is positive, symmetric, and takes values
between zero and one.

This information-theoretical distance measure, however, does not take into
account the physico-chemical similarities between amino acids. In the present
context, it is natural to focus our attention on the amino acid interactivity
$h(a)$, which determines the predicted site-specific distributions. Predicted
distributions are defined imposing the value of their first moment, so it is
natural to define the $D_{h^2}$ distance as the squared difference between the
second moments
\begin{equation}
\begin{split}
D_{h^2}(P,Q) & =
\left\{ \frac{1}{20} \sum_a h^2(a) \left[P(a)-Q(a)\right] \right\}^2 \\
& \equiv \left( [h^2]_P - [h^2]_Q \right)^2 .
\end{split}
\end{equation}

We average the distance measures between observed and predicted distributions
for different values of the structural parameter $c_i/\left< c \right>$. In the
case of the $D_{h^2}$ distance, we normalize this measure in order to reduce
its dependence on the scale of the hydrophobicity parameters. We choose the
normalization
\begin{equation}
D_{h^2}(P,Q) = \frac{\left< \left([h^2_i]_P - [h^2_i]_Q\right)^2 \right>_i}
{\left< [h^2_i]_P \right>_i \, \left< [h^2_i]_Q \right>_i} ,
\end{equation}
where angular brackets define the average over positions $i$.

\subsection{Substitution matrix}

The procedure described by Henikoff and Henikoff (1992) for calculating
substitution matrices starts by generating a multiple sequence alignment and
counting the amino acid distribution at each position,
$\pi_i^{\mathrm{ali}}(a)$. For large number of aligned positions, the logarithm
of odds matrix that they define can be approximated as
\begin{equation}\label{eq:substmat}
S_{ab} = \log\left[
\frac{\left< \pi_i^{\mathrm{ali}}(a) \, \pi_i^{\mathrm{ali}}(b) \right>_i}
{\left< \pi_i^{\mathrm{ali}}(a) \right>_i \, \left< \pi_i^{\mathrm{ali}}(b) \right>_i}
\right] ,
\end{equation}
where the angular brackets indicate an average over aligned positions $i$.

Using Eq.~(\ref{eq:substmat}), we calculate the substitution matrix generated
by aligning positions of proteins in the PDB having the same value of the
normalized PE, $c_i/\left< c \right>$, in an interval of width $0.1$.

\subsection{Normalized matrix decomposition}

Usually, the eigenvectors ${\mathbf{v}}^{(\alpha)}$ of a $N\times N$
symmetric matrix $M_{ij}$ are normalized such that their squared
components sum to one, $\sum_i \left(v_i^{(\alpha)}\right)^2 = 1$.
In this case, the spectral decomposition of the matrix can be expressed as
\begin{equation}
M_{ij} = \sum_{\alpha} \lambda_{\alpha} \, v^{(\alpha)}_i \, v^{(\alpha)}_j .
\end{equation}
The relative importance of each eigenvector, however, does not only depend on
the corresponding eigenvalue $\lambda_{\alpha}$, but also on the average
eigenvector component. To make this dependence more explicit, it is convenient
to transform the eigenvectors in such a way that their average component is
zero and their variance is one, {\it i.e.} by defining new vectors
$\eta_i^{(\alpha)} = \left(v^{(\alpha)}_i - \left< v^{(\alpha)} \right>\right)
/ \sqrt{N^{-1} - \left< v^{(\alpha)}\right>^2}$.
The spectral decomposition now reads
\begin{equation}
\begin{split}
M_{ij} = \frac{1}{N} \sum_{\alpha} \lambda_{\alpha}
\bigg[ & w_{\alpha} +
\sqrt{w_{\alpha} - w_{\alpha}^2} \,
\left( \eta_i^{(\alpha)} + \eta_j^{(\alpha)} \right) + \\
& \left( 1-w_{\alpha} \right) \, \eta_i^{(\alpha)} \eta_j^{(\alpha)}
\bigg] ,
\end{split}
\end{equation}
where $w_{\alpha} = N \, \left< v^{(\alpha)} \right>^2$. These satisfy
the normalization $\sum_{\alpha} w_{\alpha} = 1$. The first term in the
square brackets is the same for all pairs $i$ and $j$. The second term is
proportional to the sum of the normalized eigenvector components, and the
third term is proportional to their product. Therefore, we define two
combinations of the eigenvalues and the mean eigenvector components as
\begin{subequations}
\begin{equation}
\ell^{\mathrm{sum}}_{\alpha} =
\frac{\lambda_{\alpha} \, \sqrt{w_{\alpha} - w_{\alpha}^2}}
{\sum_{\beta} \lambda_{\beta} \, w_{\beta}}
\end{equation}
and
\begin{equation}\label{eq:ell-prod}
\ell^{\mathrm{prod}}_{\alpha} = \frac{\lambda_{\alpha} \, (1-w_{\alpha})}
{\sum_{\beta} \lambda_{\beta} \, w_{\beta}} .
\end{equation}
\end{subequations}


\section{Results and Discussion}

\subsection{Interactivity and folding thermodynamics}

It is well known that hydrophobicity is one of the main determinants of
protein thermodynamics. Gromiha et al.\ (1999) showed that properties
reflecting hydrophobicity correlate with the stability of buried mutations.
We tested the ability of various hydrophobicity scales to predict the average
effect of single amino acid mutations (Bastolla, unpublished).

We use a database of more than $10^3$ mutants, selected by Guerois et al.\
(2002). We average the change in folding free energy for each amino acid
substitution present in this database with more than four examples. These
average changes are correlated with the difference of hydrophobicity parameters
for various hydropathy scales. The rationale for testing these correlations is
that, from Eq.~(\ref{eq:hydro}), it follows that the difference between the
effective unfolding free energy of a protein mutated at position $i$ and the
wild-type is proportional to
$\Delta\Delta G \propto \left[ h(A_i^{\mathrm{wt}}) -
h(A_i^{\mathrm{mut}}) \right] \, \sum_j C_{ij} \, h(A_j)$.

\begin{table*}
\begin{tabular}{|c|c|c|c|c|c|c|c|c|c|}
\hline
hydrophobicity scale & $C(h,\Delta\Delta G)$ & $C(h,\alpha)$ &
$C(h,\Delta G/N)$ & $C(h,\mathrm{GC}_{12})$ & $C(h,\mathrm{genome\, size})$ &
$C(h,s_{25})$ & $C(h,s_{\mathrm{Blosum}})$ & $C(h,s_{\mathrm{PE}})$ &
$C(h,\mathrm{PE})$ \\
\hline
KD  & $0.43$ & $-0.37$ & $0.22$ & $-0.14$ & $-0.16$ & $-0.76$ & $-0.82$ & $0.74$ & $0.990$ \\
L76 & $0.53$ & $-0.35$ & $0.35$ & $-0.20$ & $-0.23$ & $-0.82$ & $-0.82$ & $0.77$ & $0.992$ \\
R88 & $0.55$ & $-0.39$ & $0.32$ & $-0.32$ & $-0.27$ & $-0.78$ & $-0.82$ & $0.74$ & $0.993$ \\
WW  & $0.63$ & $-0.46$ & $0.49$ & $-0.34$ & $-0.35$ & $-0.85$ & $-0.86$ & $0.81$ & $0.994$ \\
G98 & $0.66$ & $-0.61$ & $0.54$ & $-0.48$ & $-0.48$ & $-0.87$ & $-0.92$ & $0.83$ & $0.993$ \\
MP  & $0.67$ & $-0.45$ & $0.67$ & $-0.32$ & $-0.38$ & $-0.90$ & $-0.91$ & $0.89$ & $0.990$ \\
AV  & $0.72$ & $-0.58$ & $0.58$ & $-0.40$ & $-0.41$ & $-0.95$ & $-0.96$ & $0.93$ & $0.994$ \\
FP  & $0.77$ & $-0.52$ & $0.57$ & $-0.42$ & $-0.42$ & $-0.92$ & $-0.94$ & $0.90$ & $0.994$ \\
Z04 & $0.80$ & $-0.67$ & $0.71$ & $-0.47$ & $-0.50$ & $-0.96$ & $-0.96$ & $0.97$ & $0.994$ \\
IH  & $0.78$ & $-0.68$ & $0.83$ & $-0.59$ & $-0.55$ & $-0.94$ & $-0.94$ & $0.94$ & $0.994$ \\
CH  & $0.87$ & $-0.72$ & $0.83$ & $-0.72$ & $-0.61$ & $-0.97$ & $-0.95$ & $0.98$ & $0.995$ \\
\hline
\end{tabular}
\caption{Correlation coefficients calculated for various hydrophobicity
scales. Column~2: Changes in hydrophobicity and average change in
unfolding free energy upon mutation; Columns~3 to 6: Average
hydrophobicity of proteins in bacterial genomes and: Average normalized
energy gap (column~3), predicted unfolding free energy (column~4),
GC content in first plus second codon positions (column~5), and genome size
(column~6);
Column 7 to 9: Hydrophobicity parameters and the main eigenvector of three
substitution matrices: column~7: Matrix derived by Kinjo and Nishikawa
(2004) from aligned proteins with 25\% sequence identity (other matriceby
the same authors give essentially the same results); column~8:
The Blosum matrix (Henikoff and Henikoff, 1992); column~9: The matrix
obtained from PE alignments (this work).
Column 10: Correlation between the PE and the average HP for each PE class.}
\label{tab:hydro}
\end{table*}

\begin{figure}[t]
\begin{center}
\includegraphics[width=7.cm]{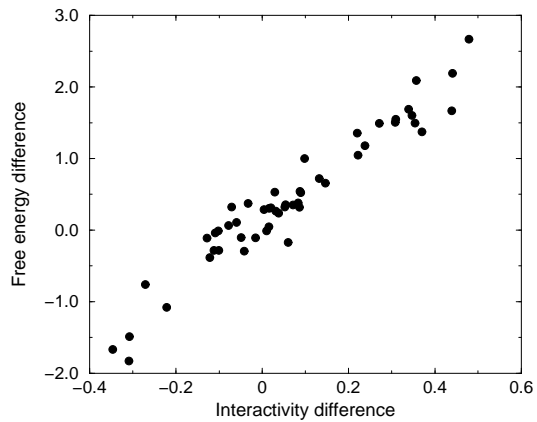}
\end{center}
\caption{Comparison between the difference of interactivity parameters
and the average free energy difference upon mutation for amino acid pairs
for which sufficient mutations have been studied.
The correlation coefficient is $R=0.87$.}
\label{fig:mut_hydro}
\end{figure}

All hydropathy scales that we use are correlated with the average changes in
the folding free energy. The strongest correlation is found for the improved
interactivity parameters CH, represented in Fig.~\ref{fig:mut_hydro}.
They provide better predictions than all the other parameters that we
examined (see Table \ref{tab:hydro}).

The influence of hydrophobicity on protein stability is complex. A more
interactive protein tends to have larger unfolding free energy, so that it is
more stable with respect to unfolding. However, alternative compact
configurations also have lower free energies, so that the normalized energy
gap is smaller and the protein may be less stable with respect to misfolding.

This qualitative picture is supported by a computational study of folding
thermodynamics in families of orthologous proteins with the same structure and
function expressed in different prokaryotic organisms (Bastolla et al.,
2004a). That study found a positive correlation between the mean
hydrophobicity and the predicted unfolding free energy, and a negative
correlation between hydrophobicity and the normalized energy gap. The
correlations hold qualitatively for all the hydropathy scales examined in
this study. The strongest
correlations are found for the optimized interactivity parameters CH, whereas
the weakest ones are found for the KD parameters (see Table~\ref{tab:hydro}).

A strong influence of hydrophobicity on protein folding thermodynamics was also
found in a recent statistical analysis of experimental results: Proteins which
are only weakly hydrophobic are unable to fold, i.e.\ they are `natively
unfolded' (Uversky, 2002a), whereas strongly hydrophobic proteins are often
characterized by folding intermediates (Uversky, 2002b), which in some cases
slow down the folding process and increase the risk of misfolding.
Interestingly, such very hydrophobic proteins are found more frequently in
obligatorily intracellular bacteria (Bastolla et al., 2004a), whose effective
population size is small due to the bottlenecks during transmission to new
hosts, so that natural selection is less effective in their evolution (Ohta,
1976). These bacteria have a very high expression level of molecular
chaperones, i.e.\ proteins that assist the folding of other proteins and
reduce the negative effects of misfolding (Hartl and Hayer-Hartl, 2002).
Therefore, it is possible that the large hydrophobicity observed in these
proteins is, at least in part, a consequence of the reduced efficiency of
natural selection, which has to be compensated
through overexpression of chaperones (Fares et al., 2004).

Possibly as a consequence of this trade-off on folding thermodynamics (Bastolla
et al., 2004a), the mean hydrophobicity of globular proteins has a rather
narrow distribution. However, for the same reason, small variations of
hydrophobicity close to the maximum of the distribution are expected to have
only a weak effect on protein folding efficiency, since they tend to be
favorable for one property but unfavorable for the other property. This might
be the reason why the GC content at first and second codon positions, which
strongly influences the interactivity of orthologous proteins (see below), is
strongly correlated with the GC content at the third, often synonymous, codon
position (Sueoka, 1961; Bernardi and Bernardi, 1986; Lobry, 1997), which, on a
genome-wide basis, is thought to reflect the mutational bias.

\subsection{Interactivity and genomic quantities}

Two important features in bacterial genomes, genome size and mean GC content
of protein coding genes, have a negative correlation with the average
interactivity of proteins expressed in those genomes, as compared with the
orthologous proteins expressed in other organisms (Bastolla et al., 2004a).
Small bacterial genomes with low GC genes (two properties that tend to go
together in prokaryotic organisms) lead to proteins of higher hydrophobicity.
We find that the optimized interactivity parameters CH have the strongest
correlations among all hydropathy scales that we considered, whereas the KD
scale has lowest and non-significant correlations (see Table~\ref{tab:hydro}).

The relationship between hydrophobicity and GC content is due to the structure
of the genetic code. As noted by several authors, Thymine in the second codon
position is almost always associated with hydrophobic amino acids. In this
respect, the low correlation found between the KD hydropathy scale and both
genomic and thermodynamic quantities is explained by the fact that this scale
attributes small or negative hydrophobicity to aromatic amino acids (Phe,
Trp, Tyr), which have large interactivity values and are AT rich in their
codons.

The relationship between hydrophobicity and prokaryotic genome size can be
attributed in part to mutation, since bacteria with small genomes tend to have
mutational bias towards AT, and in part to selection. Indeed, bacteria with
small genomes usually have an intracellular lifestyle, which implies reduced
effective populations and reduced efficiency of natural selection. This
interpretation, which we mentioned above, is supported by the observation
that the GC content at the first and second codon positions appears to be more
influenced by the mutationally driven GC content at third position in low GC
intracellular bacteria than in free living bacteria with intermediate GC in
their large genomes, indicating that negative selection is less efficient
despite the fact that the effects of very low GC on protein folding
thermodynamics are expected to be severe (Bastolla et al., 2004a).

\subsection{Optimal Hydrophobicity Profile}

Several analytic insights into protein folding and evolution can be derived
from Eq.~(\ref{eq:hydro}). First, the contact matrix whose PE is parallel
to the HP minimizes Eq.~(\ref{eq:hydro}) for a fixed value of the principal
eigenvalue. However, usually there is no protein structure that
satisfies the constraints of chain connectivity, excluded volume, and
hydrogen bonding and possesses the optimal contact matrix, so that
minimization of Eq.~(\ref{eq:hydro}) in the space of allowed protein
structures can not be performed analytically.
Second, we can define the optimal HP of a contact matrix,
${\mathbf{h}}_{\mathrm{opt}}$, as the HP that minimizes the approximate
effective free energy, Eq.~(\ref{eq:hydro}), for fixed first and second
moment of the hydrophobicity vector, $\left< h \right> = N^{-1} \sum_i h(A_i)$
and $\left<h^2\right> = N^{-1} \sum_i h(A_i)^2$. It is easy to see that the
optimal HP so defined, is very strongly correlated
with the PE (Bastolla et al., 2004b).

The condition on the mean hydrophobicity, $\left< h \right>$, is imposed
because, if a sequence is highly hydrophobic, not only the native structure
but also alternative compact structures will have favorable hydrophobic
energy. Selection to maintain a large normalized energy gap is therefore
expected to limit the value of $\left< h \right>$,
and indeed this quantity is narrowly distributed in globular proteins.
The optimal HP constitutes an analytic solution to the sequence design
problem with energy function given by Eq.~(\ref{eq:hydro}), and an
approximate solution to sequence design with the full contact energy
function, Eq.~(\ref{eq:energy}).

In the SCN evolutionary model, attempted mutations are accepted when the
effective free energy and the normalized energy gap exceed predefined
thresholds. Therefore, we do not expect the optimal HP to be ever observed
during evolution, but we do expect thermodynamically stable sequences
compatible with a given fold to have HPs distributed around the optimal one,
and therefore correlated with the corresponding PE.

The above prediction is in agreement with simulations of the SCN model.
The PE of a protein fold is significantly correlated with the HPs of its
individual SCN sequences (the mean correlation coefficient is $0.45$,
averaged over seven proteins and hundred thousands simulated sequences per
protein), and very strongly correlated with the average HP of all its
sequences. The result of this average is indicated by the symbol
$\left[ h_i\right]_{\mathrm{evol}}$, since it represents an effective
evolutionary average. Its mean correlation coefficient with the PE is
$0.96$, averaged over the same seven folds (Bastolla et al., 2004b).

Thus we can almost recover the optimal HP through an evolutionary average of
the HPs compatible with the protein fold.
Assuming that the correlation coefficient between the PE and
the average HP is exactly one, as in the case of the SCN model, we can
predict average HP as (Porto et al., 2004b)
\begin{equation}\label{eq:hevol}
\left[ h_i \right]_{\mathrm{evol}} \equiv \sum_{ \{ a \} } \pi_i(a) \, h(a) =
 A \left( c_i/\left< c \right> - 1 \right) + B \, ,
\end{equation}
where the sum over $\{ a \}$ is over all amino acids, and
\begin{subequations}
\begin{equation}\label{eq:hevol_A}
A = \sqrt{\frac{\left< \left[ h \right]_{\mathrm{evol}}^2 \right> -
 \left< \left[ h \right]_{\mathrm{evol}} \right>^2}
 {\left( \left< c^2 \right> - \left< c \right>^2 \right)/\left< c \right>^2}}
\end{equation}
and
\begin{equation}
B = \left< \left[ h \right]_{\mathrm{evol}} \right> .
\end{equation}
\end{subequations}

For protein families represented in the PFAM
(Bateman et al., 2000) and in the FSSP (Holm and Sander, 1996) databases
the PE is significantly correlated with the HP of individual sequences.
In this case, however, the correlation between the PE components and
$\left[ h_i\right]_{\mathrm{evol}}$ is considerably weaker: The mean
correlation coefficient is $0.57$ for PFAM families and $0.58$ for FSSP
families (Bastolla et al., 2004b). This weaker correlation is not unexpected,
since functional conservation is not accounted for in the SCN model, and
the effective energy function on which the interactivity is based is
only an approximation. In addition, the number of sequences used to calculate
the average HP of PFAM and FSSP families is much smaller than the number of
SCN sequences considered. When $\left[ h_i\right]_{\mathrm{evol}}$ is
computed from only a few hundred SCN sequences, as in PFAM or FSSP families,
the correlation with the PE also drops considerably.

\begin{figure}[t]
\begin{center}
\includegraphics[width=7.cm]{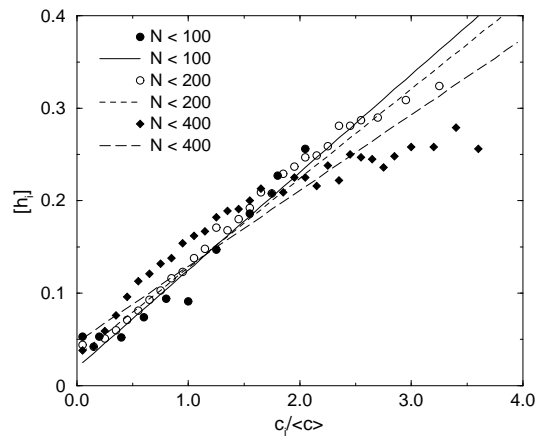}
\end{center}
\caption{Site-specific mean hydrophobicities as a function of the
corresponding normalized PE components for proteins of three different length
classes. The lines indicate the theoretical predictions,
Eq.~(\protect\ref{eq:hevol}).}
\label{fig:h_ave}
\end{figure}

We overcome these limitations by considering structural
alignments of PDB proteins based on the PE. This eliminates conservation
due to functional constraints and increases the sample size with
respect to sequence based alignments. In our prediction, the normalized PE
component at site $i$, $c_i/\left< c \right>$, determines the average
hydrophobicity at that site.
Therefore, we average together hydrophobicities at sites with similar
normalized PE components in a narrow range (see Materials and Methods).
The average HPs correlate very strongly with the PE, with correlation
coefficients very close to one for all hydrophobicity scales (see
Table~\ref{tab:hydro}). The average values are compared in
Fig.~\ref{fig:h_ave} to the values predicted through Eq.~(\ref{eq:hevol}) .
It is apparent from this figure that the prediction must be corrected for
high PE components, where the hydrophobicity values are close to the
maximum of the hydropathy scale.

Fig.~\ref{fig:h_ave} also shows that the slope of the predicted lines,
expressing the dependence of $\left[ h_i \right]_{\mathrm{evol}}$ on the PE
component, becomes smaller for longer proteins. According to
Eqs.~(\ref{eq:hevol},\ref{eq:hevol_A}), this happens because the standard
deviation (and also the mean) of the hydrophobicity profile decreases for
longer proteins.
>From Fig.~\ref{fig:h_ave} we also notice that the discrepancy between
predicted and observed $\left[ h_i \right]_{\mathrm{evol}}$ increases for
longer proteins. Consistently, we observe that longer proteins have a weaker
correlation between their PEs and HPs. These results suggest that the
selection for proper hydrophobicity is less stringent for longer proteins
(Porto et al., 2004b), consistently with the observation that longer
proteins have more native interactions per residue, and these tend to be
weaker and less optimized with respect to non-native interactions
(Bastolla and Demetrius, 2004).

\begin{figure}[t]
\begin{center}
\includegraphics[width=7.cm]{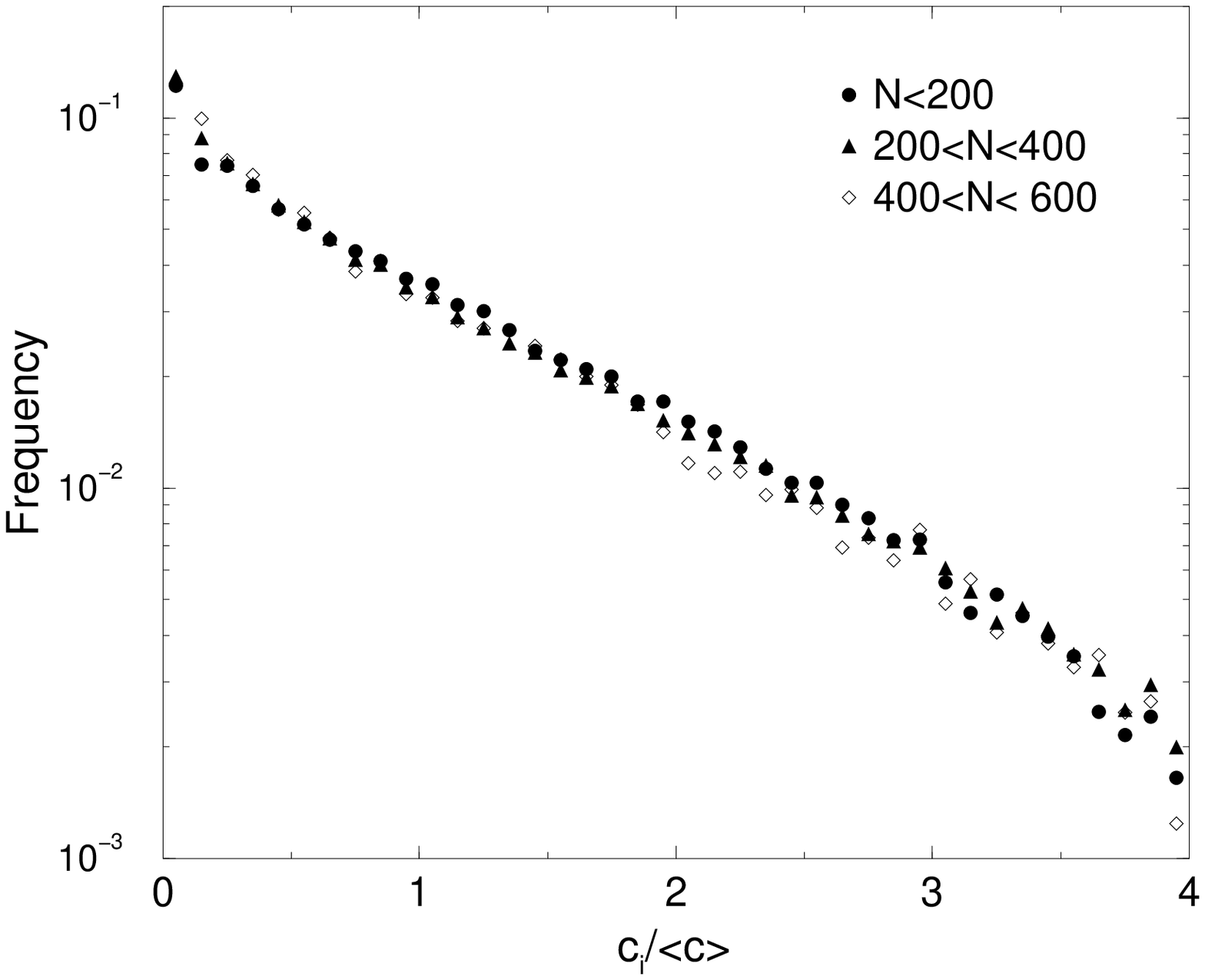}
\end{center}
\caption{Distribution of the normalized PE components,
$c_i/\left< c \right>$.}
\label{fig:P_c}
\end{figure}

The normalized PE components, $c_i/\left< c \right>$, are distributed
exponentially, as shown in Fig.~\ref{fig:P_c}. Therefore, larger PE components
are less frequently found in protein structures.

\subsection{Site-specific amino acid distributions}

The PE and the HP also allow site-specific amino acid distributions to be
predicted.
We hypothesize that the mean hydrophobicity, Eq.~(\ref{eq:hevol}), is the only
condition that the site-specific amino acid distributions have to fulfill.
Consequently, we assume that these distributions maximize the entropy for a
fixed value of $\left[ h_i \right]_{\mathrm{evol}}$. It immediately follows
that the resulting distributions have the form
\begin{equation}
\pi_i(a)\propto \exp\left[ -\beta_i \, h(a) \right] ,
\end{equation}
where the site-specific Boltzmann parameters (`inverse temperature')
$\beta_i$ determine the width of the distribution and can be obtained
inverting Eq.~(\ref{eq:hevol}).
These distributions so obtained are in very good agreement with the
site-specific amino acid distributions obtained by simulating the SCN model,
as well as with empirical distributions obtained from the PDB (Porto, 2004b).

Site-specific amino acid distributions with a Boltzmann form,
similar to those obtained above, have been used in other studies
of protein evolution. Koshi and Goldstein (1998) and Koshi et al.\ (1999)
assumed Boltzmann distributions of physico-chemical amino acid properties,
and fitted their parameters from sequence alignments using the Maximum
Likelihood method.
Dokholyan et al.\ (2001; 2002) obtained similar distributions as a
mean-field approximation of a model of protein design and evolution.
Also in this case the parameters of the distributions were obtained
from protein alignments. A main difference between these previous models
and our approach is that the latter allows to compute analytically the
Boltzmann parameters, without the need of a fitting on a database.

Despite the good agreement with observations, our approach is neglecting
a very important element, namely the mutation process acting at the DNA level.
Protein evolution takes place by mutation, which acts on individual proteins,
and by fixation of a mutant protein, which is a population-level process. We
assume here that the time scale for the appearance of new mutations is longer
than the time scale for their fixation of the population, which is the case
when the product of the effective population size $M$ times the mutation
frequency $\mu$ is smaller than one. This assumption is appropriate for
animal populations, but it is incorrect for RNA-virus
populations, and possibly for some very large bacterial populations.

We represent protein evolution as an effective stochastic process with a
transition matrix
\begin{equation}
T(a,b) = P_{\mu}(a,b) \, P_{\mathrm{fix}}(a,b)
\end{equation}
for a substitution from $a$ to $b \neq a$. The first factor represents the
mutation process and the second one represents the neutral fixation of
mutations that conserve thermodynamic stability. Our results for the SCN model
show that, for what concerns the stationary distribution, the fixation term can
be written as
\begin{equation}
P_{\mathrm{fix}}(a,b) =
\min\left\{ 1, \exp\left( -\beta_i \, [h(b)-h(a)] \right) \right\} ,
\label{eq-fix}
\end{equation}
where the Boltzmann parameter $\beta_i$ takes the value that fulfills
Eq.~(\ref{eq:hevol}). Notice that, the larger is the absolute value of
$\beta_i$, the larger is the fraction of mutations which are eliminated by
negative selection for protein stability and the larger is the mutational load.

The stationary distribution of the complete transition matrix has the form
$\pi(a,\beta)\propto \exp[-\beta \, h(a)] \, w_{\beta}(a)$, where
$w_{\beta}(a)$ satisfies the equations
\begin{equation}\label{eq:trans}
\begin{split}
0 = \sum_{a\neq b} & \min\left\{ \exp\left[ -\beta \, h(b) \right],
\exp\left[ -\beta \, h(a) \right] \right\} \times \\
& \left[ w_{\beta}(a) \, P_{\mu}(a,b) - w_{\beta}(b) \,
P_{\mu}(b,a) \right] ,
\end{split}
\end{equation}
for all final amino acid states $b$. If the mutation matrix satisfies the
detailed balance equation, $w(a) \, P_{\mu}(a,b) = w(b) \, P_{\mu}(b,a)$, then
the stationary distribution of the mutation plus fixation process has the form
\begin{equation}\label{eq:w}
\pi_i(a) = \frac{w(a)\exp\left[ -\beta_i \, h(a) \right])}
{\sum_{a^{\prime}} w(a^{\prime})\exp\left[ -\beta_i \, h(a^{\prime}) \right]} ,
\end{equation}
where $w(a)$ is the stationary distribution of the mutation process, which is
also the stationary distribution of the protein evolution process at sites
where $\beta_i$ equals zero (no rejection of mutations).

We use the simplifying hypothesis of Eq.~(\ref{eq:w}) and adopt three different
models for obtaining the weights $w(a)$. These models are scored by comparing
predicted and observed distributions through the Jensen-Shannon (JS) and $h^2$
distance measures, $D_{\mathrm{JS}}$ and $D_{h^2}$ (see Materials and Methods).
The optimized (CH) and diagonalized interactivity scales (IH) provide the best
matches between observed and predicted distributions in terms of JS distance in
almost all studied cases. The next best match is provided by the buriability
(Z04) scale. In the following, the results presented refer to the CH
interactivity scale.

We consider for comparison a model with $w(a)\equiv 1$, which corresponds to
$P_{\mu}(a,b) = 1/20$. This is the mutational model that we adopt in the SCN
simulations. Surprisingly, this simple model provides already rather good
predictions for amino acid distributions (Porto et al., 2004b). The average JS
divergence between predicted and observed frequencies in the PDB is
$D_{\mathrm{JS}} = 0.0294$ (see Table~\ref{tab:w}).

\begin{figure}[t]
\begin{center}
\includegraphics[width=7.cm]{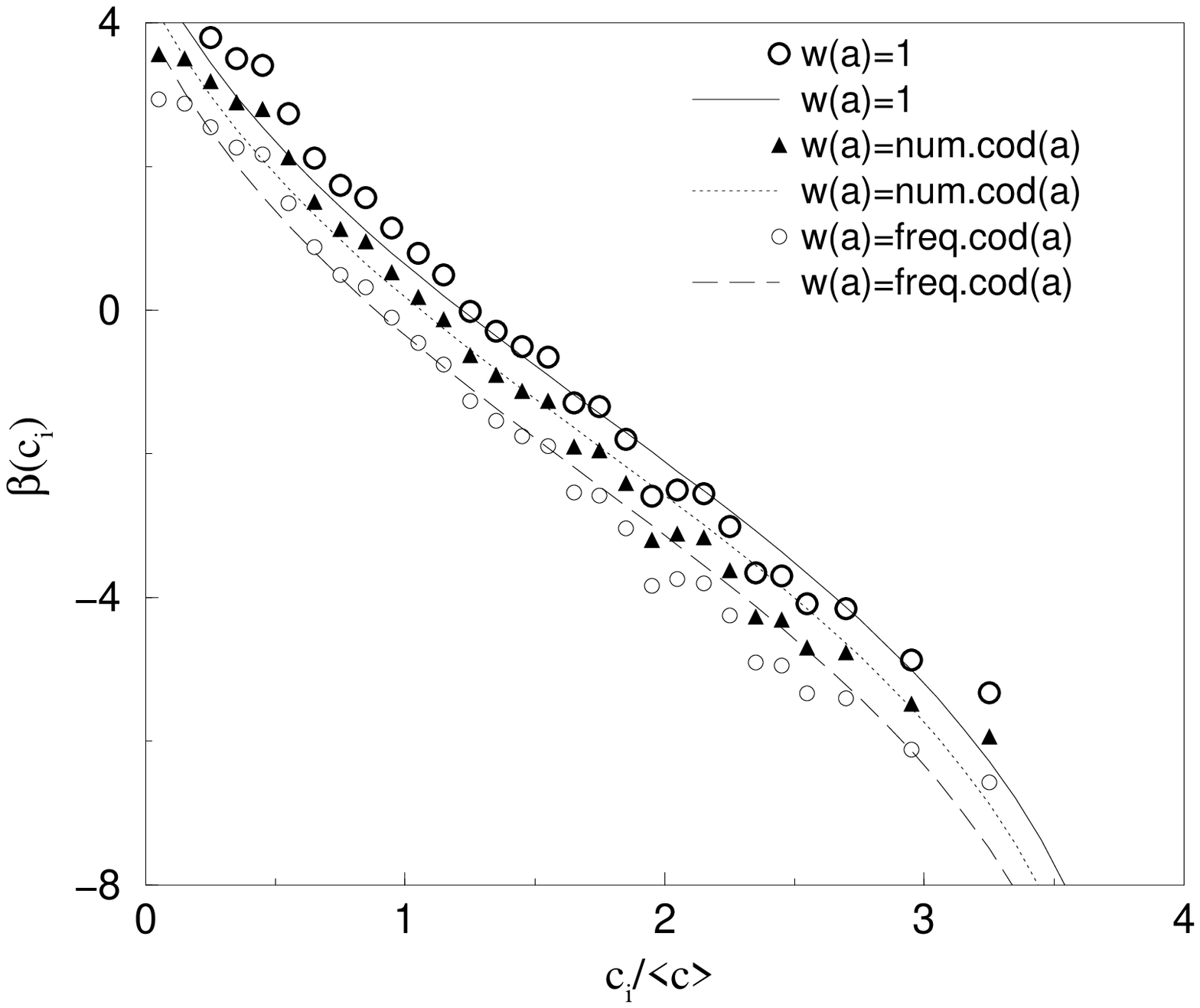}
\includegraphics[width=7.cm]{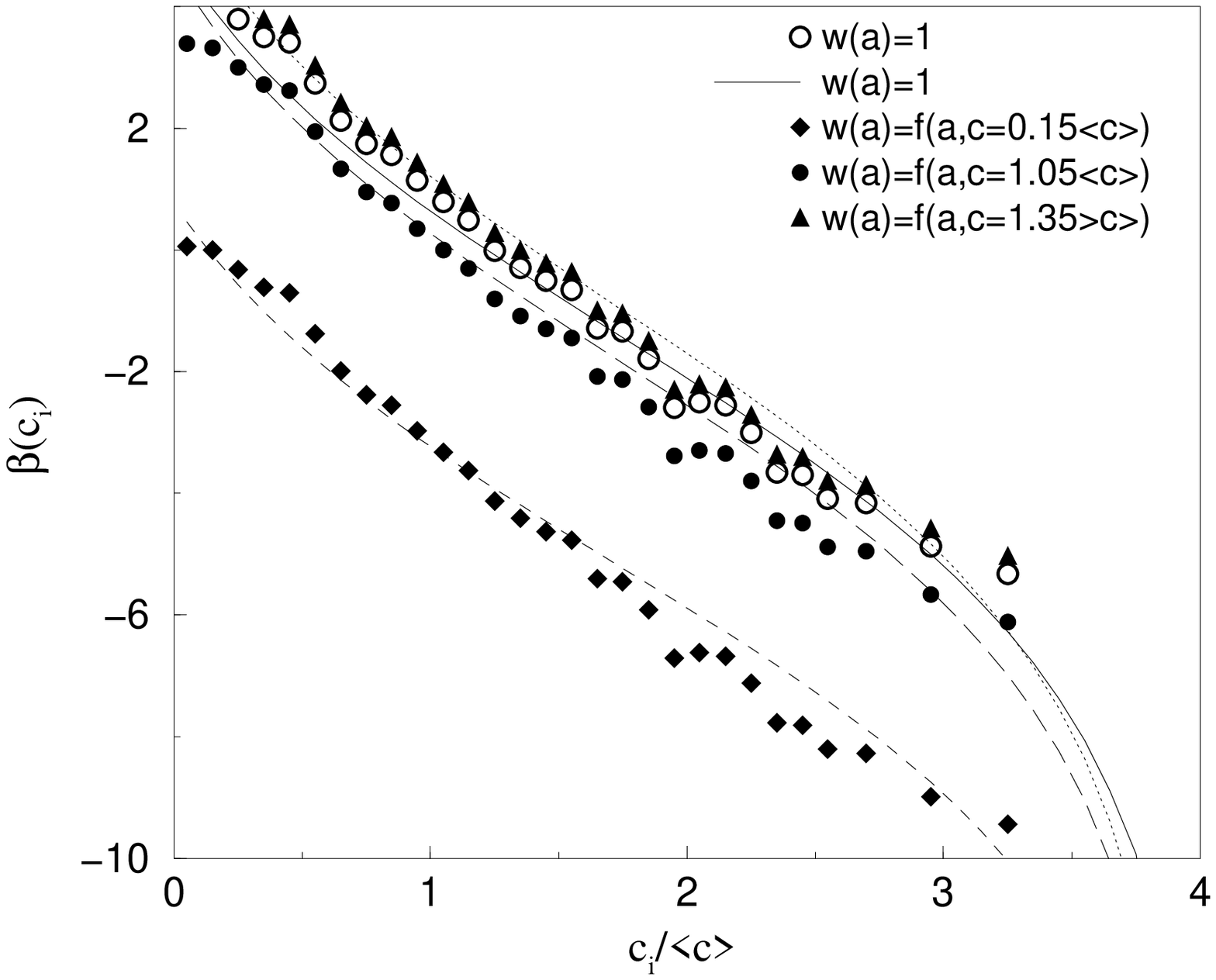}
\end{center}
\caption{Boltzmann parameters of site-specific distributions for different
choices of $w(a)$. Upper panel: $w(a)$ from number of codons and optimal codon
frequency. Lower panel: $w(a)=f(a,c)$ at $c/\left< c \right> = 0.15$, $1.05$,
and $1.35$.}
\label{fig:beta_w}
\end{figure}

In order to formulate an improved model, we choose at first $w(a)$ to be
proportional to the number of codons coding for the amino acid $a$, see
Fig.~\ref{fig:beta_w}a. This choice improves by almost 40\% the
$D_{\mathrm{JS}}$ distance, but it has little effect on the $D_{h^2}$
distance.

Secondly, we consider that not only the number of codons, but also the
nucleotide frequency influences the amino acid frequency. We adopt therefore
the expression
\begin{equation}
w(a)\propto \sum_{\mathrm{codons}(a)} f(n_1) \, f(n_2) \, f(n_3) ,
\end{equation}
where $f(n)$ is the frequency of the four nucleotides $\mathrm{A}$,
$\mathrm{T}$, $\mathrm{G}$, and $\mathrm{C}$. The previous ansatz based on the
number of codons is equivalent to assuming identical frequencies for the four
nucleotides. The
nucleotide frequencies providing the optimal match are found by enumeration of
all combinations in a reasonable range, with discretization step of $0.005$.
For the IH interactivity parameters, the optimal frequencies are $\mathrm{T} =
0.240$, $\mathrm{C} = 0.195$, $\mathrm{A} = 0.310$, and $\mathrm{G} = 0.255$.
These frequencies are close to observed nucleotide frequencies in protein
coding genes, and they do not change much when we use other hydropathy scales,
or the D$_{h^2}$ measure instead of D$_{\mathrm{JS}}$. The improvement in
D$_{\mathrm{JS}}$ is a further 30\%, but the improvement in D$_{h^2}$ is
modest.

The third approach considers that, for $\beta=0$, the expected amino acid
frequencies coincide with the weights $w(a)$. We therefore obtain $w(a)$
parameters from the frequencies observed at various values of $c/\left< c
\right>$, where we effectively set $\beta=0$, see Fig.~\ref{fig:beta_w}b.
Note that this approach requires to obtain $19$
additional parameters from the data and it does not correspond to any
mutational model. Therefore, the fact that it better fits the data is not
significant, and we show these results only for comparison. In this case
the improvement of the D$_{\mathrm{JS}}$ measure is dramatic, as expected, for
$c/\left< c \right>$ in the range between zero and two. The D$_{h^2}$ measure
improves only modestly, and it becomes worse for $c/\left< c \right> > 1.2$.
The combination of the two criteria and the comparison of $\beta$ values from
observed and predicted distributions favour weights derived at $c/\left< c
\right>$ between zero and one. Notice however that, if $c/\left< c \right>$
is much
smaller than one, meaning that we fit the weight at positions with small PE
component, the required $\beta$ parameters are much more negative. This fact
can be interpreted as a strong selective pressure acting on positions with
large PE component.

\begin{table}[t]
\begin{tabular}{|l|c|c|c|c|}
\hline
$w(a)$ & D$_{\mathrm{JS}}$ & D$_{h^2}$ & D$_{\beta}$ & $\left<| \beta_i |\right>$ \\
\hline
constant                             & 0.0294 & 0.338 & 0.128 & 0.416 \\
number of codons                     & 0.0247 & 0.282 & 0.136 & 0.388 \\
optimal codon frequency              & 0.0172 & 0.271 & 0.129 & 0.362 \\
uniform codon mutations              & 0.0256 & 0.285 & 0.256 & 0.393 \\
frequency at $c=1.05\left< c\right>$ & 0.0087 & 0.265 & 0.177 & 0.380 \\
\hline
\end{tabular}
\caption{Distances and average absolute Boltzmann parameter
for distributions obtained for various models of $w(a)$.}
\label{tab:w}
\end{table}

Lastly, we solve Eq.~(\ref{eq:trans}) numerically for a mutation matrix
obtained by considering all nucleotide mutations as equiprobable. This process
has a stationary distribution that is almost proportional to the number of
codons of each amino acid. Accordingly, its prediction coincides almost
perfectly with the one obtained by choosing $w(a)$ proportional to the
number of codons, even if detailed balance is not assumed. This result
justifies the approximation of considering matrices that satisfy detailed
balance. Results corresponding to various models are summarized in
Table~\ref{tab:w}.

\subsection{Rejection rate}

Using the mutation matrix $P_{\mu}(a,b)$ and the fixation probability
Eq.~(\ref{eq-fix}), we can compute the equilibrium fraction of mutations which
are eliminated by negative selection at site $i$ with Boltzmann parameter
$\beta$ (we omit the index $i$ for simplicity),
\begin{equation}
\begin{split}
P_{\mathrm{rej}}(\beta) =
\sum_{a=1}^{20}
\sum_{\genfrac{}{}{0pt}{}{\scriptstyle b \text{ with}}
 {\scriptstyle\beta \, h(a) < \beta \, h(b)}} \!\!\!\!
& w_{\beta}(a) \, P_{\mu}(a,b) \times\\
& \left(\mathrm{e}^{-\beta \, h(a)} - \mathrm{e}^{-\beta \, h(b)} \right)\, .
\end{split}
\end{equation}
The computation of this quantity requires to know the mutation matrix,
whereas our assumption of detailed balance only requires to specify the
stationary nucleotides frequencies under no selection ($\beta=0$).
Therefore, we evalutate the rejection rate through the average of the
absolute value of the Boltzmann parameters $\left<|\beta_i|\right>$.

Among the models that we considered, $\left<|\beta_i|\right>$ is lowest for
the more elaborate model taking into account nucleotide frequencies. Among all
possible nucleotide frequencies, the lowest value of $\left<|\beta_i|\right>$ is
$0.348$, and it is attained at nucleotide frequencies close to those giving the
optimal fit, $\mathrm{T} = 0.200$, $\mathrm{A} = 0.320$, $\mathrm{C} = 0.180$,
and $\mathrm{G} = 0.300$. Even lower values of $\left< |\beta_i| \right>$ and
better fits can be obtained deriving the $w(a)$ from the amino acid frequencies
observed at specific values of the PE component (the minimum value of $\left<
|\beta_i| \right>$ is $0.328$, attained when the $w(a)$ coincide with the
frequencies observed at $c_i = 0.75 \left< c \right>$), but these values are
ad-hoc since they are not the result of a mutational model. These results
suggest that mutation frequencies are set to almost optimize the mutational
load.

\subsection{Interactivity and substitution matrices}

Kinjo and Nishikawa (2004) recently calculated substitution matrices derived
from protein alignments with various values of the sequence identity of the
aligned proteins. Qualitatively, these substitution matrices measure the
tendency of two residue of types $a$ and $b$ to co-occur at aligned sites.
Kinjo and Nishikawa noticed that the eigenvector corresponding to the largest
positive eigenvalue of the substitution matrices is correlated with the
hydrophobicity, and that the corresponding eigenvalue increases with
decreasing sequence alignment, corresponding to longer evolutionary time.

We examine the substitution matrices calculated by Kinjo and Nishikawa for
various sequence identities, and we consider for comparison the matrix
Blosum62 (Henikoff and Henikoff, 1992). We also obtain a new substitution
matrix from the alignment of PDB structures based on the normalized PE
components, as described in Materials and Methods. This latter substitution
matrix can also be predicted analytically using the formulas for site-specific
distributions given above. Its non-diagonal elements are correlated with the
corresponding elements of the Blosum62 matrix, with $R=0.66$.
However its diagonal elements are much smaller than the corresponding Blosum
elements, since all 20 amino acids can be found with fairly large frequencies
at each aligned position.

We calculate the correlation of the eigenvector corresponding to the largest
positive eigenvalue of these substitution matrices with the hydrophobicity
scales described in Materials and Methods. For almost all matrices, the
largest correlation (with $R = -0.95$ to $-0.97$) is found for the optimized
interactivity parameters, but the correlation is very high also for other
scales, see Table~\ref{tab:hydro}.

\begin{figure}[t]
\begin{center}
\includegraphics[width=7.cm]{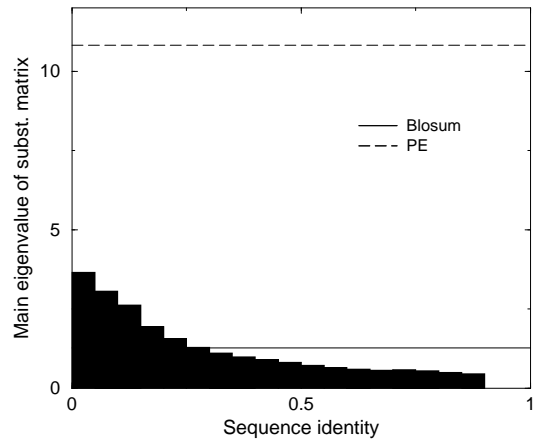}
\end{center}
\caption{Normalized eigenvalue corresponding to the hydrophobicity
contribution to substitution matrices as a function of the sequence identity
used to derive the matrix. The Blosum matrix and the matrix derived from
PE-based alignment are also shown.}
\label{fig:subst_mat}
\end{figure}

We then computed the normalized eigenvalue of this eigenvector,
Eq.~(\ref{eq:ell-prod}). From its behavior as a function of sequence identity
(see Fig.~\ref{fig:subst_mat}) one sees that the contribution of the
hydrophobicity to the substitution matrices becomes more important for lower
identities of the aligned proteins, {\it i.e.} for increased evolutionary
time. The Blosum matrix corresponds to an intermediate value, whereas the
matrix obtained from aligned PDB structure is completely dominated by
hydrophobicity, as expected from Eq.~(\ref{eq:substmat}). This matrix
corresponds to the limit of very long evolutionary time, where all sequence
similarity is lost, apart for that originated by the common structural
environment expressed by the PE. Therefore,
this study of substitution matrices further supports our results on the
relationship between the PE and sequence evolution.


\section{Conclusions}

It is well known that the hydrophobicity is a major determinant of protein
stability and evolution. We have shown that, among several hydropathy scales,
the optimized interactivity scale (Bastolla et al., 2004b) is the one best
correlated with various thermodynamic and genomic quantities.

The influence of the hydrophobicity on protein stability has two sides: On
the one hand, the more hydrophobic a protein is, the more stable is with
respect to
unfolding; on the other hand it is less stable with respect to misfolding, so
that a careful tuning of the hydrophobic content is needed (Bastolla et al.,
2004a). Close to the optimal hydrophobicity, small variations have complex
effects that might reduce their impact on fitness. This is probably the reason
why the mutational bias is able to influence hydrophobicity and hence protein
folding thermodynamics. Large variations of the mean hydrophobicity, however,
are not tolerated in the course of evolution.

In order to achieve folding stability, we have predicted that the
hydrophobicity should be distributed within the protein sequence in a
site-specific manner, which is mainly influenced by a site-specific indicator
-- the principal eigenvector of the contact matrix (PE). From the hypothesis
that the correlation between the hydrophobicity and the PE is the only
condition imposed by folding stability on sequence evolution, we predicted
analytically site-specific amino acid distributions to be of Boltzmann type,
in agreement with simulations of the SCN model of protein evolution and
with structural alignments of sites with equivalent PE taken from the PDB.

Finally, we have discussed how simple mutational models improve the fit
between predicted and observed site-specific distributions. Interestingly, the
model that provides the best fit is also one with almost the lowest rate of
rejected mutations. This observation is consistent with the suggestions that
error correction systems
might have evolved to minimize the impact of mutations on protein stability.


\flushleft

\end{document}